\documentclass[reprint, amsmath,amssymb,prb]{revtex4-2}
\usepackage{graphicx}
\usepackage{dcolumn}
\setcitestyle{numbers,square}
\usepackage{braket}
\usepackage{graphicx}
\usepackage{dcolumn}
\usepackage{bm}
\usepackage[normalem]{ulem} 
\usepackage{color}
\usepackage{hyperref}
\usepackage{xspace}
\usepackage{footnote}
\usepackage[title,titletoc,toc]{appendix}
\hypersetup{colorlinks,citecolor=blue,linkcolor=blue,urlcolor=blue}
\usepackage[section]{placeins}
\newcommand{\bip}{BiP$_3$}
\newcommand{\mbip}{m-BiP$_3$}

\begin{document}
\title{Unveiling the electronic properties of BiP$_3$
triphosphide from bulk to graphene-based heterostructure by
first-principles calculations }

\author{Dominike P. de Andrade Deus}
\email{dominike@iftm.edu.br}
\affiliation{Instituto Federal de Educa\c{c}\~ao, Ci\^encia e Tecnologia do Tri\^angulo Mineiro, Uberaba, C. P. 1020, 38064-190, Minas Gerais, Brazil}

\author{Igor S. S. de Oliveira}
\email{igor.oliveira@ufla.br}
\affiliation{Departamento de F\'isica, Universidade Federal de Lavras, C.P. 3037, 37203-202, Lavras, MG, Brazil}

\author{Roberto Hiroki Miwa}
\email{hiroki@ufu.br}
\affiliation{Universidade Federal de Uberlândia, Av. João Naves de Ávila, 2121 - MG, 38400-902, Brazil}

\author{Erika N. Lima}
\email{erika@ufr.edu.br}
\affiliation{Instituto de Ciências Exatas e Naturais, Universidade Federal de Rondonópolis, Av. dos Estudantes, 5055 - Cidade Universitária, Rondonópolis - MT, 78736-900, Brazil}
\affiliation{Instituto de Física, Universidade Federal de Mato Grosso, Cuiaba, Mato Grosso 78060-900,
Brazil}        

\date{\today}
  
\begin{abstract}
\vspace{3mm}
\begin{center}
 {\bf ABSTRACT}
\end{center}

Triphosphides, with a chemical formula of XP$_3$ (X is a group IIIA, IVA, or VA element), have recently attracted much attention due to their great potential in several applications. Here, using density functional theory calculations, we describe for the first time the structural and electronic properties of the bulk bismuth triphosphide (BiP$_3$). Phonon spectra and molecular dynamics simulations confirm that the 3D crystal of BiP$_3$ is a metal thermodynamically stable with no bandgap. Unlike the bulk, the mono-, bi-, tri-, and tetra-layers of BiP$_3$ are semiconductors with a bandgap ranging from 1.4 to 0.06 eV. However, stackings with more than five layers exhibit metallic behavior equal to the bulk. The results show that quantum confinement is a powerful tool for tuning the electronic properties of BiP$_3$ triphosphide, making it suitable for technological applications. Building on this, the electronic properties of van der Waals heterostructure constructed by graphene (G) and the \bip~monolayer (m-\bip) were investigated. Our results show that the Dirac cone in graphene remains intact in this heterostructure. At the equilibrium interlayer distance, the G/m-BiP$_3$ forms an n-type contact with a Schottky barrier height of 0.5 eV. It is worth noting that the SHB in the G/m-BiP$_3$ heterostructure can be adjusted by changing the interlayer distance or applying a transverse electric field. Thus, we show that few-layers \bip~is an interesting material for realizing nanoelectronic and optoelectronic devices and is an excellent option for designing Schottky nanoelectronic devices.


\end{abstract}
\maketitle

\section{Introduction}
The emergence of graphene\cite{geim1} as a two-dimensional material consisting of a single layer of carbon atoms arranged in a hexagonal lattice has sparked significant scientific interest in recent years. Its exceptional mechanical strength\cite{geim2,Novoselov2012}, electrical conductivity\cite{grapheneconductive}, and thermal conductivity\cite{Balandin2008} make it a material of immense potential for various fields, including electronics\cite{graphenedevice0,graphenedevice1,graphenedevice2}, energy storage\cite{Raccichini2015}, and biomedical engineering\cite{bioengineeringLianzhou2015}. The unique properties of graphene have opened up new avenues for scientific exploration and advanced material development. As a result, researchers worldwide are studying graphene to unlock its full potential and develop novel applications for this extraordinary material. In light of its remarkable properties, graphene has attracted considerable attention in the field of materials science, particularly in the development of heterostructures\cite{grapheneheterostructure1,grapheneheterostructure2,grapheneheterostructure3,grapheneheterostructure4} composed of graphene and other materials, such as metallic or semiconductor substrates\cite{graphenesemiconductor1,graphenesemiconductor2} as well as other two-dimensional (2D) materials like boron nitride\cite{grapheneBN1,grapheneBN2} or transition metal dichalcogenides\cite{grapheneTMD1,grapheneTMD2,grapheneTMD3}. 

Also, more than 30 years ago, several compounds metal triphosphides belonging to the family of II-V and III-V, such as SnP$_3$\cite{triphosphide1snp3}, GaP$_3$\cite{triphosphide8GeP3}, GeP$_3$\cite{triphosphide2Gep3GeP5}, SnP$_5$\cite{triphosphide2Gep3GeP5}, and InP$_3$\cite{triphosphide3InP3}, have been successfully synthesized. These compounds were obtained by reacting different elements under high-pressure and high-temperature conditions, and their crystal structures were determined by single-crystal X-ray diffraction or powder-diffraction intensity data. Indeed, theoretical calculations have been done on 2D triphosphide materials. The $\alpha$-BP$_3$\cite{triphosphide3BP3} crystal, a two-dimensional material composed of phosphorus and boron atoms, is energetically more stable than another allotrope of phosphorene and exhibits high electron mobility ($\approx$ 4.6$\times$ 10$^{4}$cm$^2$V$^{-1}$s$^{-1}$) in both monolayer and bilayer forms. Similarly, SnP$_3$\cite{triphosphide4SnP3} has been predicted to have low cleavage energies, tunable band gaps, high carrier mobilities in both mono- and bilayer, and significant optical absorption due to the existence of van-Hove singularities in the electronic density of states. Meanwhile, SnP$_3$\cite{triphosphide5SnP3} has been found to possess a direct band gap and ultrahigh carrier mobility comparable to that of monolayer phosphorene, and its band gap can be tuned over a wide range by controlling the number of stacked layers. Additionally, CaP$_3$\cite{triphosphide6CaP3} in a two-dimensional form can absorb visible light across the entire spectrum. Finally, GeP$_3$\cite{triphosphide7GeP3} mono- and bilayer show low indirect band gaps  ($\approx$ 0.50 eV), high carrier mobility has potential applications in photovoltaics, and magnetic properties when mixed with foreign transition metal atoms\cite{dominike}. These findings suggest that 2D triphosphide materials have significant potential as functional materials for future nanoelectronic and optoelectronic applications. Most recently, research focused on the potential of 2D triphosphide materials, particularly in heterostructures, has prompted further exploration into the characteristics of XP3 compounds when deposited on various substrates, especially graphene, as a promising platform for its high mechanical and electrical properties\cite{graphene3XP3,graphene2XP3,graphene1XP3}.

In this work, we have investigated the structural
and electronic properties of the bulk, few layers, and monolayer
forms of the BiP$_3$ triphosphide. In the bulk, the most stable
arrangement presents an ABC stacking. Phonon and Molecular
Dynamics (MD) calculations confirm that the bulk material is
thermodynamically stable. Interestingly, when the effect of
spin-orbit coupling (SOC) is considered in the calculations, the
material transitioned from a semiconductor to a metallic system.
On the other hand, in the case of the \bip~monolayer,  the
semiconductor character is maintained both in the absence and
presence of SOC, and it has an indirect bandgap of 1.41 eV.
Moreover, it presents a low exfoliation energy. We also explored
the structural and electronic properties of the heterostructure
formed by the combination of graphene (G) and monolayer
\bip~(m-\bip). Our investigation confirms the preservation of the
properties of both G and m-\bip~when they come into contact,
forming an n-type Schottky contact. Additionally, we show that the
Schottky barrier heights can be adjusted by modifying the distance
between layers or by applying an external electric field
perpendicular to the layers.

\section{Computational details}
Our DFT calculations were performed within the Perdew-Burke-Ernzehof generalized gradient approximation\cite{gga1}, using the projector augmented wave (PAW) potentials,\cite{paw} as implemented in the Vienna Ab-initio Simulation Package(VASP) \cite{vasp1}. The structural optimizations were done, including van der Waals corrections (vdW)\cite{grimme} until the forces on each atom were lower than 0.01 eV/\AA~ and total energies converged within a 10$^{-6}$~eV criterion. The Kohn-Sham (KS)wave functions were expanded in a plane-wave basis set with an energy cutoff ($E_\text{cut}$) of 600 eV. The sampling of the Brillouin-zone (BZ) was performed by using the Monkhorst-Pack (MP)\cite{mp} scheme with k-point meshes of 5$\times$5$\times$1 and 10$\times$10$\times$1 for structural optimization and electronic properties, respectively. Spin-orbit coupling (SOC) was included in all electronic structure calculations. Phonon-properties were conducted using the density functional perturbation theory (DFPT) method, as implemented in the PHONOPY code \cite{phonopy}.

\section{\bip~ bulk}

\subsection{Structural properties}

\begin{figure}[htb!]
    \includegraphics[width=\columnwidth]{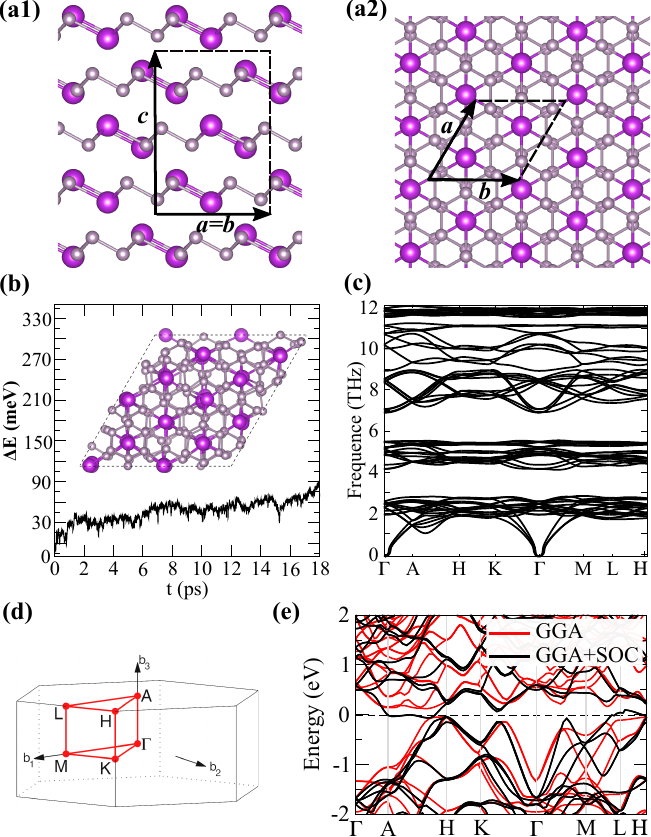}
    \caption{\label{fig:stackings_models} (a1) Side and (a2) top views of a \bip-bulk compound in the ABC-stacking. The hexagonal supercell's lattice parameters, $a$, $b$, and $c$, are marked with black dashed lines. (b) \textit{Ab-initio} molecular dynamics (AIMD) and (c) phonon dispersion for the \bip-bulk. (d) High symmetry path in the hexagonal Brillouin to the hexagonal crystal structure, and (e) energy bulk dispersion of \bip~in the ABC-stacking.}
\end{figure}

We initially investigated the structural properties to acquire a comprehensive understanding of \bip-bulk in AAA-, ABA-, and ABC-stacked layers.  We verified that the ABC-stacking [Fig. \ref{fig:stackings_models}-(a1)] showed the lowest total energy considering semi- and non-local van der Waals potentials [Fig. \ref{fig:SM1}-(b1) in Appendix-\ref{label:geometries}]. 
This is due to the strong electrostatic interaction between the
positively charged bismuth and negatively charged phosphorus ions
in the adjacent layers. In the ABC-stacking configuration, each
phosphorus atom has a bismuth atom directly above and below it,
forming a triangular lattice. This leads to the formation of a
stable crystal structure with strong interlayer bonding, resulting
in a lower total energy of the material. The AAA- and ABA-stacking
configurations have weaker interlayer bonding  (or a strong
repulsion between layers) due to the lack of a triangular lattice
and a different arrangement of bismuth and phosphorus atoms,
resulting in higher total energy states. For consistency, we used
the DFT-D2 method of Grimme for all analyses in this work, as the
lattice parameters, atomic and interlayers distances were found to
be similar for all admitted Vdw corrections, as shown in figure
Fig. \ref{fig:SM1}-(b1) of Appendix-\ref{label:geometries}. Each
bismuth (Bi) atom has three bonds with neighboring phosphorus (P)
atoms. The crystal structure of the \bip-bulk is characterized by
a lattice with parameters $a = b$ = 7.83~\AA~[Fig.
\ref{fig:stackings_models}-(a2)], and $c/a$ = 1.34~\AA, where $c$
refers to the parameter perpendicular to each single layer, which
gives rise to a rhombohedral lattice system. This system is
identified as belonging to the trigonal crystal symmetry, with
lattice described by the crystallographic symbol R3$\bar{m}$ and
point group 3$\bar{m}$. The lattice parameters indicate that the
material has a hexagonal unit supercell with a threefold axis of
symmetry. Additionally, the values of $\alpha$ = $\beta$ =
90\textdegree, and $\gamma$= 120\textdegree~ show that the crystal
structure possesses a threefold rotation axis and three
perpendicular mirror planes. In ABC-stacking, the distance between
two adjacent layers is 2.10~\AA. The interatomic distance analysis
of \bip~reveals that the separation between phosphorus and bismuth
atoms is approximately d$_{Bi-P}$ = 2.78~\AA. In contrast, the
distance between P and P is around $d_{P-P}$=2.24~\AA. In addition
to examining the ABC stacking, we conducted a comprehensive
analysis of the AAA and ABA stackings of \bip-bulk. An in-depth
understanding of these structural analyses can be found in
Appendix-\ref{label:geometries} of the Supplementary Material.

In our study, we also conducted phonon dispersion analysis [Fig. \ref{fig:stackings_models}-(b)] to investigate the vibrational properties of \bip-bulk. The results of our analysis revealed that there were no negative frequencies, indicating that the bulk structure is stable and free from any imaginary modes of vibration.  We also analyzed the dynamics and thermodynamic stability of \bip~using AIMD [Fig. \ref{fig:stackings_models}-(c)]. The findings show that the bulk structure is stable at room temperature and remained unchanged for 18 ps with minimal energy change results.  Based on the result, it can be concluded that the arrangement of \bip-bulk in an ABC stacking configuration is stable.
\subsection{Electronic properties}
To investigate the electronic properties of \bip~bulk, the initial step involved the computation of the electronic structure of the bulk phase without considering the effect of spin-orbit coupling (SOC), which revealed a semiconductor behavior characterized by an indirect bandgap of 0.17 eV [red lines in Fig. \ref{fig:stackings_models}-[(d)]. However, High-resolution ARPES experiments have recently been conducted to analyze the surface and bulk bands of Bi, and the results strongly indicate the existence of nontrivial band topology in pure Bi\cite{bismuth1Ohtsubo_2013,bismuth2PERFETTI2015,bismuth3Ito2016}.  Subsequently, the SOC was incorporated into the calculation, and the resulting electronic structure of the \bip~system became metallic [black lines in Fig. \ref{fig:stackings_models}-[(d)]. The SOC coupling in \bip-bulk creates electronic connectivity between layers. This is evident along the $\Gamma$-H and M-H directions, where the electronic properties turn metallic, indicating significant modifications to interlayer interactions. Furthermore, the HSE06 energy dispersion (Fig. \ref{fig:bulkhse} in Appendix-\ref{label:bulkhse}) reveals that the metallic states near the Fermi level in the bulk remain largely consistent, providing further validation for the theoretical findings.

\section{\bip~ monolayer}

\begin{figure}[htb!]
    \includegraphics[width=\columnwidth]{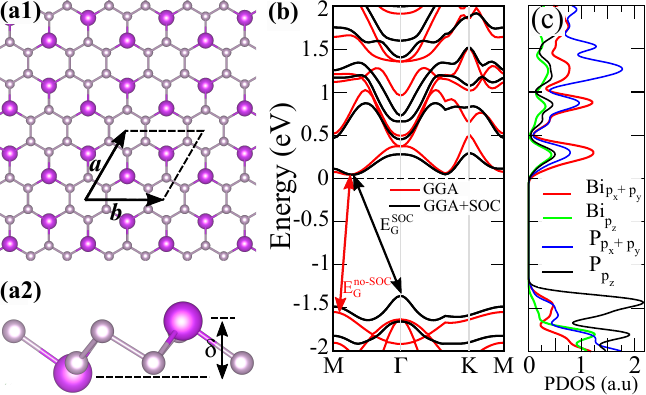}
    \caption{\label{fig:monolayer} (a1) Top and (a2) side view of m-\bip structure. (b) The band structure of m-\bip~by considering GGA (red lines) and GGA+SOC (black lines). (c) The corresponding pDOS for the p-orbitals of m-\bip~was obtained from GGA+SOC+HSE06 calculation. }
\end{figure}

We proceed with our investigation by admitting the structure
properties of the m-\bip, which has the $P-3m1$ trigonal symmetry
group and a hexagonal honeycomb lattice characterized by a
puckered atomic layer. By using phonon dispersion spectra, Such
monolayers were previously confirmed by Yi-Yuan Wu et
al.\cite{bip3phonon1} and Hong-Yao Liu et al.\cite{LIUBiP3}. We
calculated the energy required to exfoliate one or two layers of
\bip, and our findings demonstrate applicable exfoliation energies
of 1.07 and 0.96 J/$m^{2}$~for monolayer and bilayer extraction
from a six-layer (6L) \bip~slab, respectively. The
Appendix-\ref{label:exfoliation} provides details on obtaining the
exfoliation energy. These values are close to the cleavage
energies of other triphosphides, e.g., GeP$_3$ (1.14 J/$m^{2}$)
\cite{triphosphide7GeP3}, InP$_3$ (1.32 J/$m^{2}$)
\cite{inp3exfoliation}, CaP$_3$ (1.30 J/$m^{2}$)
\cite{triphosphide6CaP3}, and SnP$_3$ (0.57 J/$m^{2}$)
\cite{triphosphide5SnP3}. Regarding the structure, the Bi atom
forms three Bi-P bonds with three adjacent P atoms, while the P
atom forms one Bi-P bond and two P-P bonds. The lattice parameters
for this monolayer are optimized with $a$ = $b$ = 7.11 Å [Fig.
\ref{fig:monolayer}-(a1)]. Compared to the \bip-bulk, the
relaxation process of the m-\bip~resulted in a significant
reduction of approximately 9\% in the in-plane lattice parameters
$a$ and $b$, leading to a more pronounced puckered structure
characterized by a larger $\delta$ value ($\delta_{\rm
BiP_3-bulk}$=1.26 \AA$~\rightarrow~$ $\delta_{\rm m-BiP_3}$=1.55
\AA)[Fig. \ref{fig:monolayer}-(a2)].  However, the bonding
distances between atoms remained unchanged. %
Additionally, we have found that m-\bip~has a band structure with
an indirect gap of 1.55 eV [Fig. \ref{fig:monolayer}-(b)].  The
electronic properties of m-\bip~are greatly affected by SOC. The
band structure between non-SOC and SOC calculations shows
significant differences, with the SOC-induced spin splitting of
the bands resulting in a smaller indirect gap in the SOC
calculation than in the no-SOC ($E_{G}^{non-SOC}=$ 1.55
eV~$\rightarrow$~$E_{G}^{SOC}=$ 1.40 eV). The valence band maximum
(VBM) is at the center of the Brillouin zone, specifically at the
$\Gamma$ point, while the conduction band minimum (CBM) lies
between the $\Gamma$ and M high symmetry points. Additionally, the
bands exhibit anisotropic behavior, which varies depending on the
crystal momentum direction. Hence, SOC plays a crucial role in
comprehending the electronic structure and transport properties of
m-\bip. It is interesting to note that Hong-Yao Liu et al.
\cite{LIUBiP3} did not consider SOC in their calculations, yet
they observed that the gap energy and parabolic behavior around
the $\Gamma$ point were similar to those obtained from SOC
calculations. This implies that the electronic properties of
m-\bip~may not be solely influenced by SOC, and other factors such
as crystal structure and chemical bonding must also have an
impact. Additionally, we utilized a hybrid functional approach
(GGA+SOC+HSE06) to further investigate the electronic properties
of m-\bip, and we found out that the band gap increased from 1.40
eV to 2.00 eV [Fig. \ref{fig:hsemonolayer} in
Appendix-\ref{label:hsemonolayer}]. Moreover, the VBM and CBM
experienced similar shifts down and up due to the HSE06's
inclusion concerning the pure SOC calculation. Notably, despite
these alterations, the dispersion and band indirect gap were quite
preserved. Based on the partial density of states (pDOS) obtained
from the SOC+HSE06 calculation, it can be inferred that the states
near the Fermi level are primarily derived from P$_{p_{z}}$
orbitals in the vVBM, and Bi$_{p_{x}+p_{y}}$ and P$_{p_{x}+p_{y}}$
orbitals in the CBM [Fig. \ref{fig:monolayer}-(c)].

\section{\bip~few layers}
We thoroughly investigated the electronic and structural
characteristics of \bip~with few layers [Fig. , ranging from two
(2L) [Fig. \ref{fig:models_2l_to_8l}-(a) in
Appendix-\ref{label:models_2l_to_8l}] to eight layers (8L) [Fig.
\ref{fig:models_2l_to_8l}-(g) in
Appendix-\ref{label:models_2l_to_8l}] . We have considered
AB-stacking for the bilayer. From 3L onwards, the layers follow
ABC stacking. Our analysis of the structural properties revealed
exciting trends in the evolution of the lattice parameter from the
monolayer to multilayer regimes. We noticed that as the layer
count increased from a monolayer (7.11 \AA) to eight layers (7.43
\AA), the lattice parameter grew consistently, indicating a
gradual convergence towards the lattice parameter of the bulk
material (7.83~\AA), as can be seen in the
Table-\ref{tab:fewlayers}. Additionally, it has been observed
that the interlayer separation diminishes as the number of layers
comprising the material increases. In the context of a bilayer
configuration, the interlayer equilibrium separation is 2.63 \AA.
However, as the number of layers increases from 3L to 8L, the
interlayer distance adjacent to the surface's edges measures
approximately 2.17 \AA, whereas towards the central region of the
surface, the interlayer separation contracts to an approximate
value of 2.00 \AA. This distinctive behavior is attributed to the
amplification of van der Waals forces between neighboring layers
as the layer count ascends, consequently compelling them to draw
closer together. Furthermore, as the stacking arrangement
increases, the distance between layers decreases, approaching the
bulk configuration of \bip. In the \bip-bulk, the interlayer
distance is quantified at around 2.10 \AA. This tendency is
explained by the denser packing of layers, leading to an
interlayer spacing close to the \bip's bulk form.

Regarding the electronic properties of a few layers, we noticed that the characteristics of \bip~changed as the number of layers is altered, as shown by the band gaps values compiled in Table-\ref{tab:fewlayers}. The energy gap has decreased from 1L to 4L, suggesting a shift towards the metallic electronic behavior of the bulk. The metallic state was verified in configurations of five layers or more (5L, 6L, 7L, and 8L). Details on the band structures of \bip~ from 2L to 8L can be seen in Fig.\ref{fig:bandfewlayers} of Appendix-\ref{label:bandfewlayers}.  


\begin{table}[h!]
\begin{tabular}{ c c c }
\hline
\textbf{Layers} & \textbf{\textit{a$_{0}$}(\AA)} & \textbf{Band gap (eV)} \\ \hline
1 & 7.11 & 1.40      \\  
2 & 7.17 & 0.60      \\  
3 & 7.32 & 0.08      \\  
4 & 7.36 & 0.06      \\  
5 & 7.39 & metallic  \\  
6 & 7.40  & metallic \\ 
7 & 7.41 &  metallic  \\ 
8 & 7.43  & metallic \\  \hline
$bulk$ & 7.83  & metallic \\  \hline 
\end{tabular}
\caption{\label{tab:fewlayers} Lattice parameters (second column) and band gaps (third column) as a function of number of layers (first column).}
\end{table}



\section{Graphene/BiP$_3$ heterostructure}

\begin{figure}
    \includegraphics[width=\columnwidth]{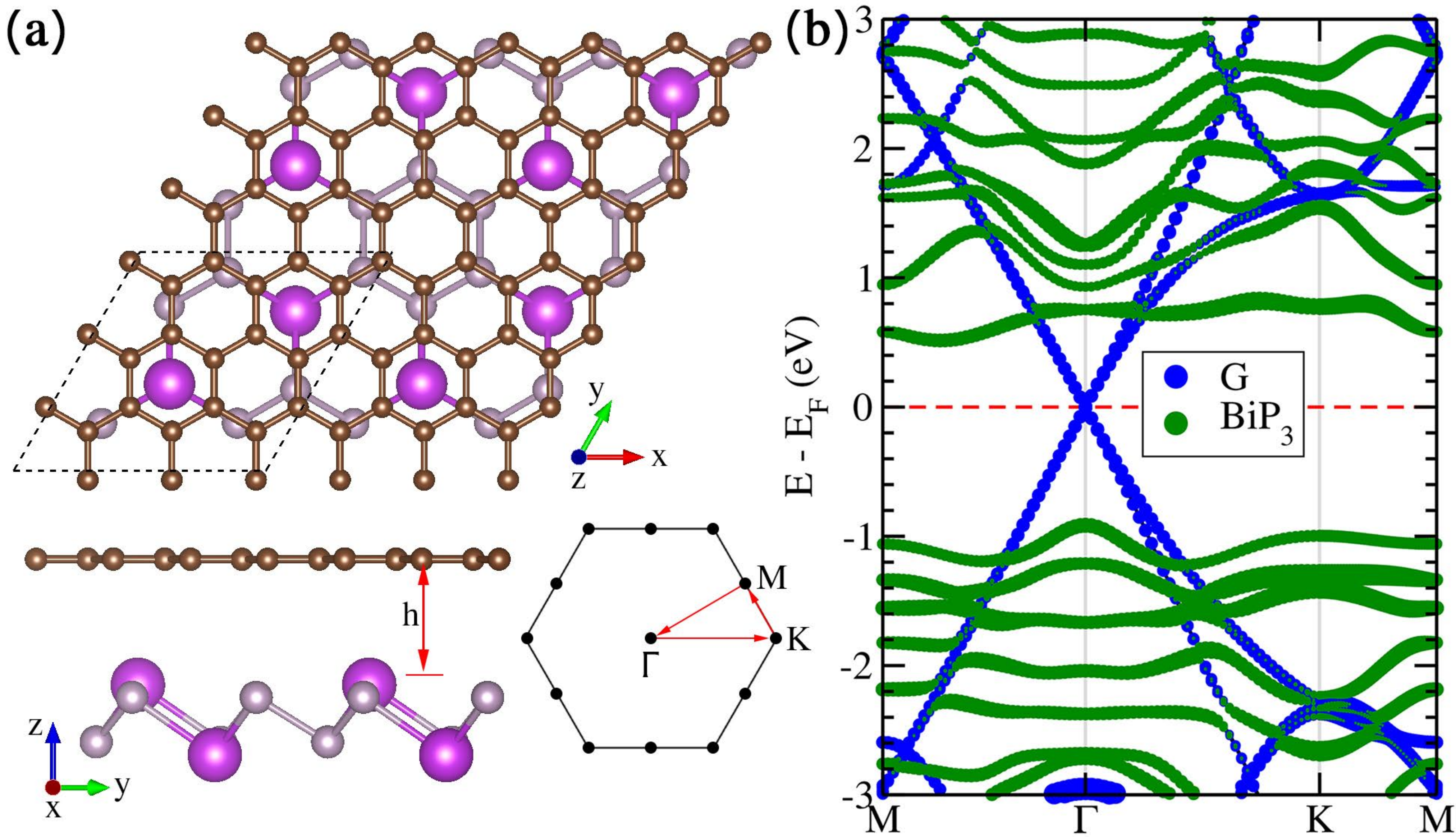}
    \caption{(a) Top and front views of the G/\mbip\ equilibrium geometry. The black hexagon represents the Brillouin zone of the G/\mbip\. (b) Projected band structure of the G-m/\bip. The line width of blue and green indicates the projected weights of the G and \mbip\ components, respectively.}
    \label{strucbandsHet} 
\end{figure}

We now investigate the electronic and structural properties of the vdW heterostructure formed by the deposition of a single layer of graphene (G) on top of the \bip~ monolayer 
(G/m-\bip). The simulation supercell consists of a ($3\times 3$) graphene and a ($1\times 1$) \bip\ cells. The pristine primitive graphene cell has a lattice parameter of $a=b=2.46$~\AA, resulting in a lattice mismatch of $\sim 3.8\%$ with the m-\bip. As the m-\bip\ band gap is sensitive to applied strains, we choose to fix the m-\bip\ lattice parameters, hence compressing the graphene layer. We have considered three highly symmetric configurations for depositing the G-layer on top of the m-\bip, as presented in Fig.\ref{fig:configsgbip3} of Appendix-\ref{label:configsgbip3}. For each arrangement, we calculate the binding energy ($E_{\rm b}$) using the following equation:
\begin{equation}
E_{\rm b}=[ E_{\rm {G}/m-BiP_3} - (E_{\rm G}+E_{\rm BiP_3}) ] / N_{\rm C},
\end{equation}
where $E_{\rm {G}/m-BiP_{3}}$, $E_{\rm G}$, and $E_{\rm BiP_3}$ represent the total energies of the G/m-\bip\ junction, pristine G, and pristine \mbip, respectively. $N_{\rm C}$ is the number of carbon atoms in the G layer. We find that the $E_{\rm b}$ process is exothermic for all simulated configurations with a similar value, with a maximum difference 
of only about 1\% between them.
The most energetically stable stacking configuration presents 
the hole of the graphene hexagonal rings on top of the upmost Bi of the 
\bip, with $E_{\rm b}=-42.32$~meV/C-atom, which is shown in Fig.~\ref{strucbandsHet}(a). 
One can typically find this $E_{\rm b}$ order in graphene vdW heterostructures \cite{graphene3XP3,grapheneheterostructure4}.
The interlayer distance $h$ between G and m-\bip\ 
[see Fig.~\ref{strucbandsHet}(a)] in the equilibrium geometry is 3.35~Å, which suggests that there are no chemical bonds between them. Furthermore, the structural properties of both G and m-\bip\ remain unaffected as they are brought closer, indicating that weak vdW forces govern the interactions between G and m-\bip.

We then proceed to examine the electronic characteristics of the G/m-\bip\ vdW heterostructure. As demonstrated earlier, including SOC is crucial for describing the electronic structure of \mbip. Therefore, SOC is taken into account for all G/\mbip\ electronic analyses. Fig.~\ref{strucbandsHet}(b) displays the projected band structure, which reveals the presence of the graphene linear Dirac-like dispersion relation around the Fermi level. The \mbip\ layer retains its semiconductor properties from the pristine monolayer, with an indirect energy gap of 1.42~eV between the VBM and CBM, which is 20~$m$eV larger than the isolated \mbip\ band gap. This increase in the band gap might be attributed to the weak interaction between \mbip\ and the G layer. These results demonstrate that the electronic properties of isolated G and \mbip\ layers are well-preserved in the G/\mbip\ heterostructure formation, indicating a small degree of interaction at the interface.

\begin{figure}
    \includegraphics[width=\columnwidth]{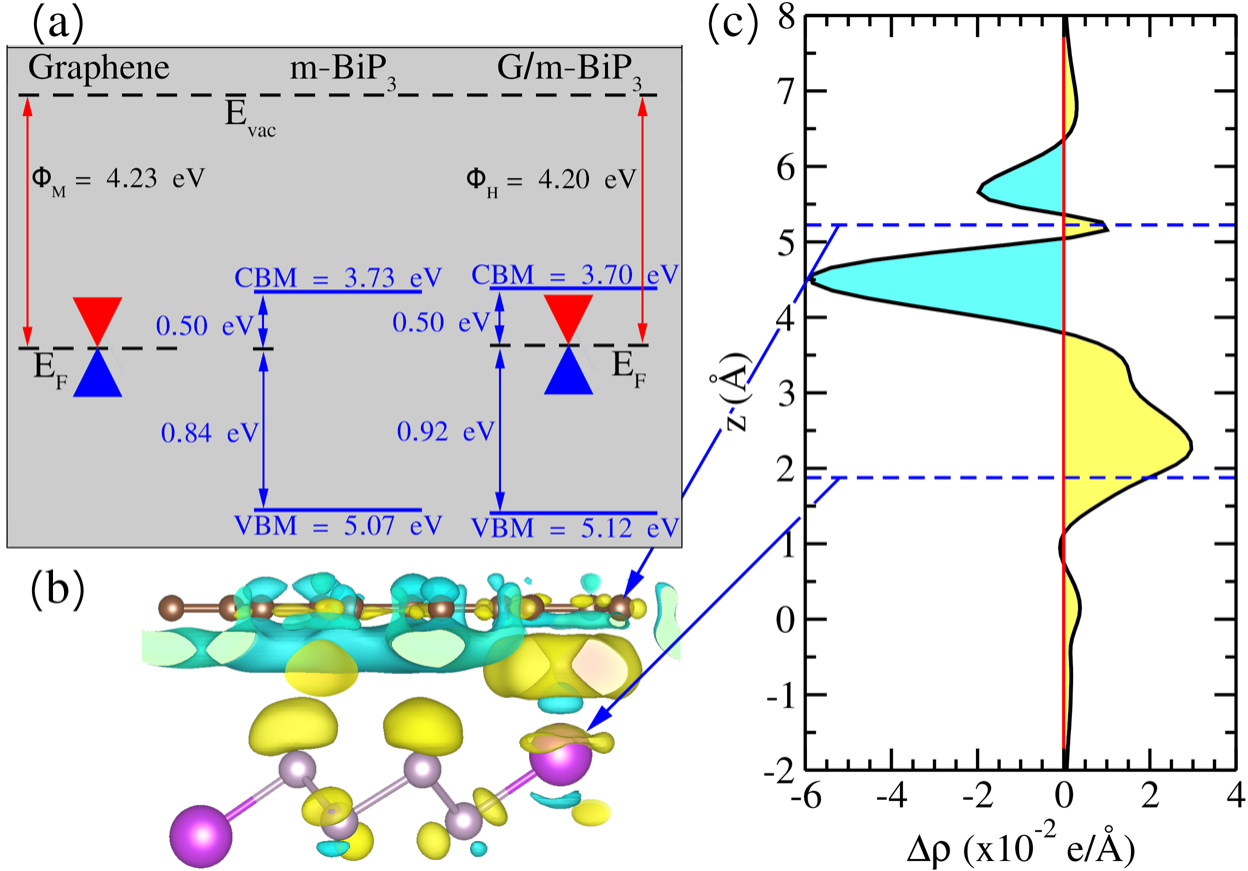}
    \caption{\label{drho} (a) Representation of the energy levels for graphene, \mbip, and G/\mbip. (b) Differential charge density of G/\mbip, (c) the plane-averaged DCD, computed along the perpendicular direction ($z$-axis).}
\end{figure}

Recently, weak interactions in certain 2D van der Waals (vdW) heterostructures have been investigated to design metal-semiconductor junctions with low chemical bond formation, chemical disorder, and defect density.  In such scenarios, the Schottky-Mott rule proves useful for estimating the Schottky barrier height (SBH) of the system.\cite{Liu2018}
The weak Fermi level pinning on metal-2D semiconductor vdW junctions allows the tuning of the SBH through the metal working function ($\Phi_{\rm M}$) modulation.\cite{PhysRevLett.114.066803, Liue1600069}
As shown in Fig.~\ref{drho}(a), we find a work function
$\Phi_{\rm M}=4.23$~eV for the free-standing graphene, 
in agreement with previous theoretical and experimental works. \cite{doi:10.1021/nl901572a,garg2014work,leenaerts2016work}
Moreover, the isolated \mbip\ CBM and VBM energy values 
are informed in Fig.~\ref{drho}(a). Taking the vacuum energy ($E_{\rm vac}$) as a reference, we note that the graphene Dirac cone lies between those values. 
The G/\mbip\ band structure [Fig.~\ref{strucbandsHet}(a)] suggests the absence (or very minimal) of Fermi level pinning at the junction. This observation suggests that the Schottky-Mott rule can be employed to determine the SBH in the G/\mbip\ heterostructure.
Thus, we define a $n$-type SBH ($\Phi_{\rm Bn}$) as the difference between the \mbip\ CBM energy and the graphene Fermi level: $\Phi_{\rm Bn} = E_{\rm CBM} - \Phi_{\rm M}$. Similarly, the $p$-type SBH is defined as $\Phi_{\rm Bp} = \Phi_{\rm M} - E_{\rm VBM}$, where $E_{\rm VBM}$ represents the VBM energy. These definitions yield $\Phi_{\rm Bn} + \Phi_{\rm Bp} \approx E_{\rm gap}$, where $E_{\rm gap}$ denotes the energy gap of \mbip.
For the equilibrium system, we find $\Phi_{\rm Bn} = 0.50$ eV and $\Phi_{\rm Bp} = 0.92$ eV, as illustrated in 
Fig.~\ref{drho}(a). 
Since the graphene Dirac cone lies closer to the \mbip\ CBM than its VBM energy level, the G/\mbip\ heterostructure constitutes a $n$-type Schottky contact, with electron conduction prevailing within the system.

In Fig.~\ref{drho}(b) we plot differential charge density, 
defined as
    \begin{equation}
        \Delta\rho(\mathbf{r}) = \rho_{\rm G/m-BiP_3}(\mathbf{r}) - \rho_{\rm G}(\mathbf{r})
                        - \rho_{\rm m-BiP_3}(\mathbf{r}),
    \end{equation}
where $\rho_{\rm G/m-BiP_3}(\mathbf{r})$ represents the electronic density for the heterostructure, $\rho_{\rm G}(\mathbf{r})$ corresponds to pristine G, and $\rho_{\rm BiP_3}(\mathbf{r})$ represents isolated \mbip. 
The obtained results reveal a charge depletion in the graphene region (blue isosurfaces), accompanied by an accumulation of charge on the \mbip\ surface (yellow isosurfaces). To further investigate the redistribution of charge within the system, we also calculate the integral of the differential charge density (DCD) over the $xy$-plane using the following equation:
\begin{equation}
\Delta\rho(z) = \int_{\rm S} \Delta\rho(\mathbf{r}),dx,dy,
\end{equation}
where $\Delta\rho(z)$ represents the integrated DCD, and S denotes the surface area of the supercell. The outcome of this calculation is depicted in Fig.~\ref{drho}(c), clearly illustrating the observed electronic charge accumulation on graphene and depletion on \mbip. By employing the Bader charge analysis method
\cite{Bader,Henkelman}, we verified an increase of 0.033$\,e$ 
($7.6\times 10^{-12}$~$e$/cm$^2$) in the G layer, coming from
the \mbip. The small amount of charge transfer suggests the formation of weak vdW interactions at the heterojunction interface.

\begin{figure}
    \includegraphics[width=\columnwidth]{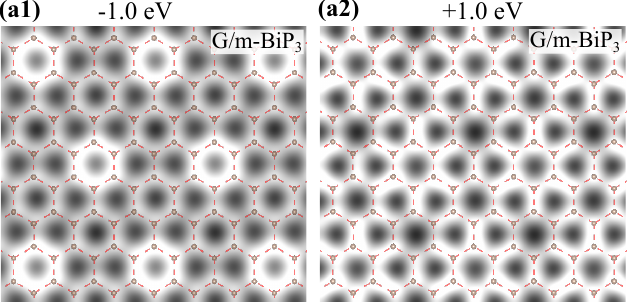}
    \caption{\label{fig:stm4} STM obtained at a constant height of 2\AA. The saturation level ranges from zero to 5$\times$10$^6$ e$^{-}$/bohr${^2}$. The red dashed lines indicate the positions of the graphene carbon hexagons.}
\end{figure}

Utilizing the theoretical Tersoff-Hamann
method\cite{tersoff}, Scanning Tunneling Microscopy (STM)
simulations were conducted to investigate the atomic-scale
features within the G/m-\bip~system. STM images were obtained at
energy levels corresponding to E$_F~\pm$ 1.0 eV, yielding the
following observations: (i) the absence of a hexagonal pattern in
both occupied [Fig.\ref{fig:stm4}-(a)] and empty
[Fig.\ref{fig:stm4}-(b)] states suggest an interaction between the
graphene layer and m-\bip, deviating from the pristine graphene's
expected hexagonal symmetry; (ii) bright spots observed in both
empty and occupied states stem from specific interactions between
the topmost bismuth of m-\bip~and carbon atoms within the graphene
layer, indicating prominent features in the STM images originate
from this Bi-C interaction (in agreement with differential charge
density in Fig.\ref{drho}-(b)); (iii) in occupied states, bright
states result from the hybridization between Bi$_{p_{z}}$ and
C$_{p_{z}}$ orbitals. This hybridization is similarly observed in
unoccupied states.
\begin{figure}
    \includegraphics[width=\columnwidth]{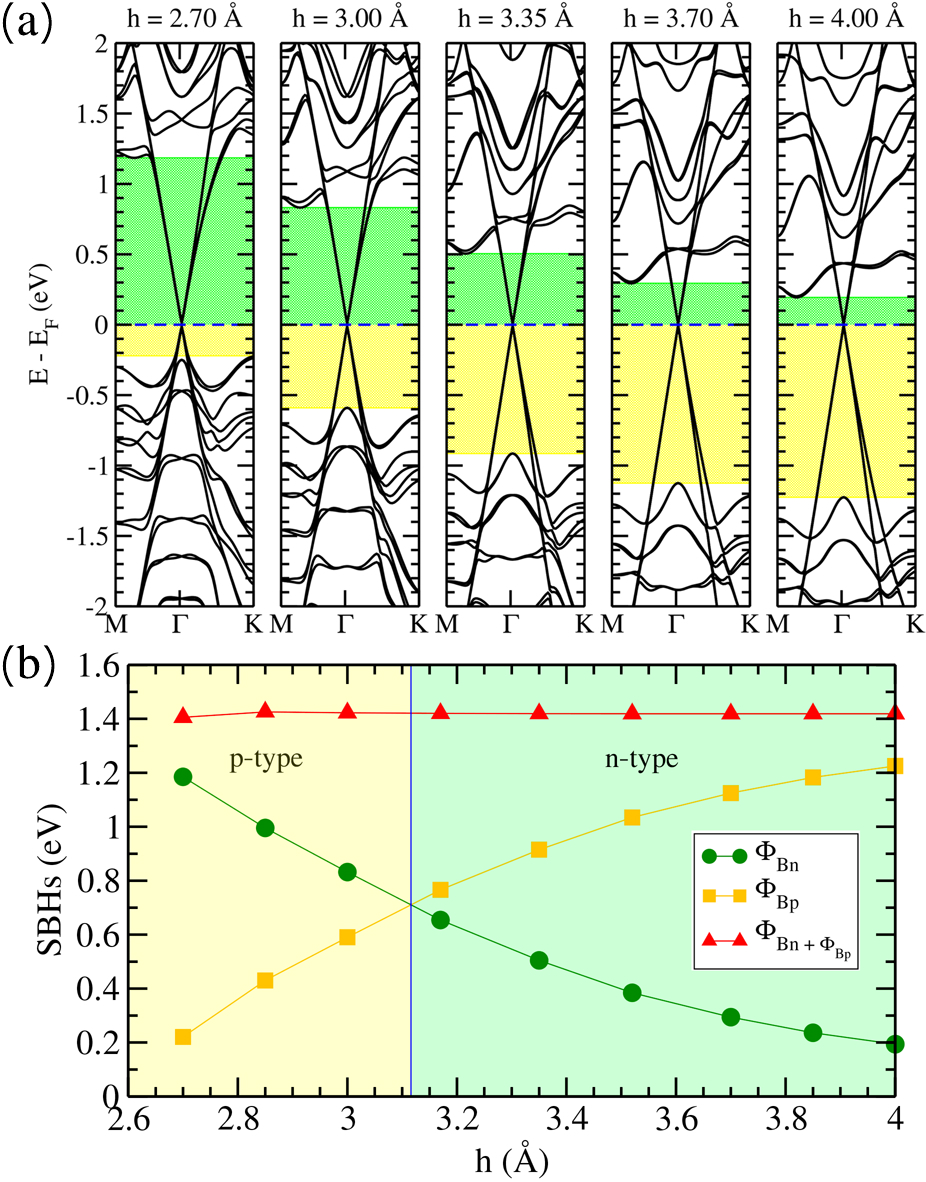}
    \caption{\label{pressure} (a) Band structures and 
    (b) Schottky barrier heights of G/\mbip\ heterostructure for various interlayer distances.}
\end{figure}

Enhancing the performance of metal-2D semiconductor vdW contacts in electronic devices has posed a significant challenge, primarily due to the difficulty in controlling the SBH. To address this, researchers have explored various approaches, including adjusting the SBH through interlayer coupling modulation. In our system, this can be accomplished by modulating the interlayer distance ($h$) between the G layer and the topmost atomic layer of \mbip. 
To examine the impact of modifying $h$ on the electronic properties of G/\mbip, we computed the band structure for various interlayer distances, as depicted in Fig.~\ref{pressure}(a). As the interlayer distance increases from 
$3.35\rightarrow 3.70\rightarrow 4.00$ Å, the graphene Dirac cone gradually shifts towards the CBM of the \mbip, while the CT decreases 
from $0.033\rightarrow 0.026\rightarrow 0.025 \,e$.
Consequently, this leads to a decrease in $\Phi_{\rm Bn}$ and an increase in $\Phi_{\rm Bp}$.
On the other hand, when reducing $h$ from 3.35 to 3.00~\AA\ ($CT=0.035\,e$)
and further to 2.70~\AA\ ($CT=0.060\,e$), there is an observed increase in $\Phi_{\rm Bn}$ and a corresponding decrease in $\Phi_{\rm Bp}$.
In Fig.~\ref{pressure}(b), we present the SBHs for different interlayer distances. As the interlayer distance ($h$) increases from its equilibrium value, the $\Phi_{\rm Bp}$ value increases, while $\Phi_{\rm Bn}$ decreases. 
Conversely, when the interlayer distance is decreased from its equilibrium value, an opposite trend is observed. At $d=3.17$~\AA, $\Phi_{\rm Bn}$ and $\Phi_{\rm Bp}$ become close to each other, and below this point, the magnitudes of the SBHs are inverted, meaning $\Phi_{\rm Bp}$ becomes smaller than $\Phi_{\rm Bn}$. Consequently, by adjusting the interlayer distance, a transition from $n$-type to $p$-type Schottky contact can be induced in the G/\mbip\ heterostructure. Throughout the process of varying $h$, the energy gap between the \mbip\ VBM and CBM remains nearly constant, as indicated by the red triangles in Fig.~\ref{pressure}(b).


\begin{figure}
    \includegraphics[width=\columnwidth]{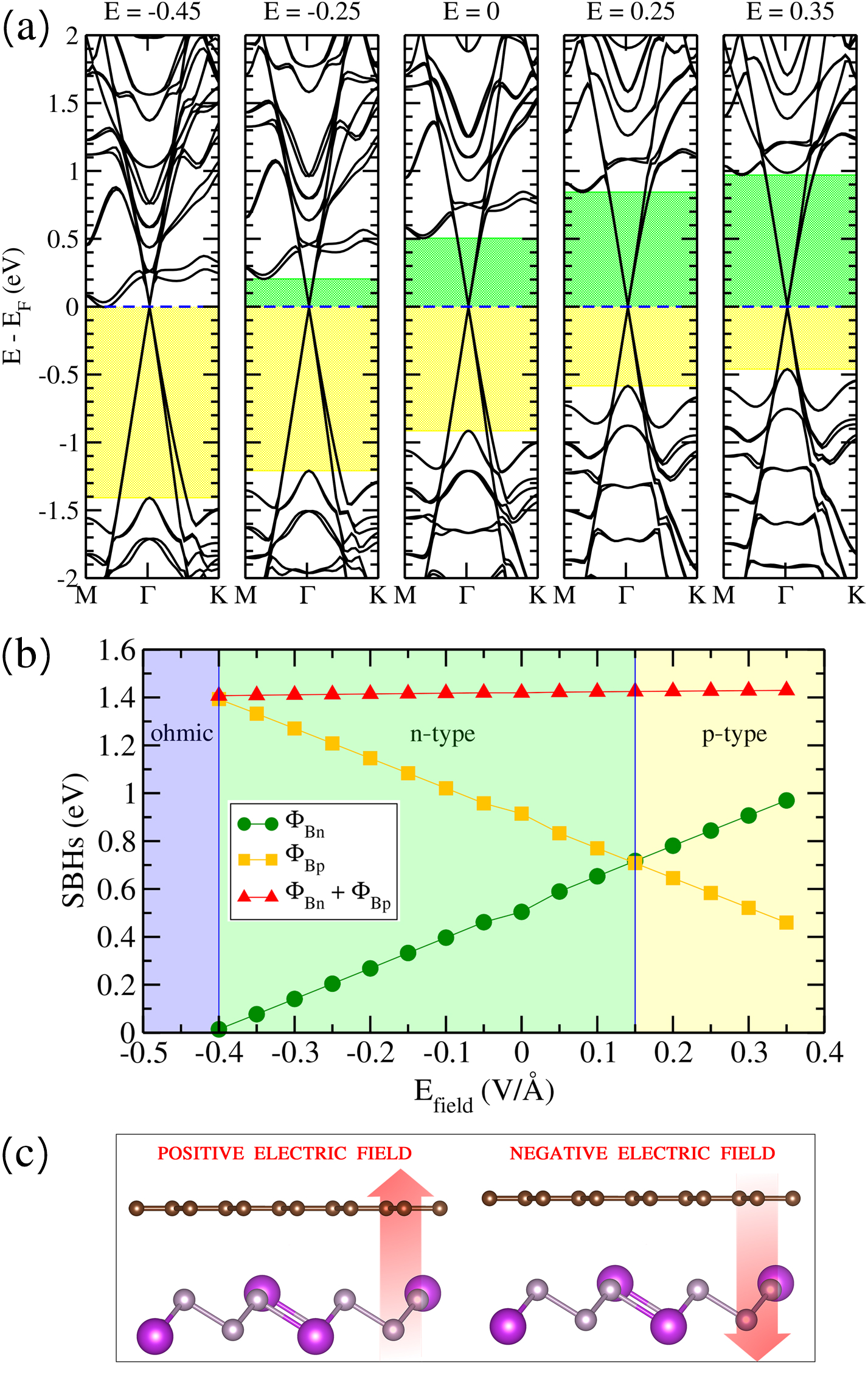}
    \caption{\label{efield} (a) Band structures and 
    (b) Schottky barrier heights of G/\mbip\ heterostructure for various applied external electric field values. In (a)
    the $E$ values are given in units of eV/\AA. (c)
    Side view of the G/\mbip\ vdW heterostructure when a transverse electric field is applied. The upward (downward) red arrow illustrates the electric field's positive (negative) direction.}
\end{figure}
Applying a perpendicular external electric field ($E_{\rm field}$) is another commonly employed method to modify the SBHs in G vdW heterostructures. 
The external transverse electric field values in our calculations range from -0.45 to 0.35~V/\AA. The positive direction of the electric field points from \bip\ to graphene, whereas the negative direction points in the opposite direction, as seen in Fig.~\ref{efield}(c). Consequently, we investigate the impact of the electric gating on the electronic properties of the G/\mbip\ heterostructure in Fig.~\ref{efield}(a) we present the band structure for selected $E_{\rm field}$ values.
For positive values ($E_{\rm field} = 0.25$ and $0.35$~V/\AA), 
we observe an increase in $\Phi_{\rm Bn}$ and
a decrease in $\Phi_{\rm Bp}$. On the other hand, for 
$E_{\rm field} = -0.25$~V/\AA\ $\Phi_{\rm Bn}$ decreases and 
$\Phi_{\rm Bp}$ increases. At $E_{\rm field} = -0.45$~V/\AA,
although $\Phi_{\rm Bp}$ continues to increase, a conduction band crosses the Fermi level, forming an ohmic contact.

Fig.~\ref{efield}(b) shows the computed SBHs for different values of $E_{\rm field}$. Positive values of $E_{\rm field}$ lead to an increase in the $\Phi_{\rm Bn}$ while $\Phi_{\rm Bp}$ decreases, both exhibiting a linear relationship. 
When $E_{\rm field}$ approaches 0.15~V/\AA, $\Phi_{\rm Bp}$ approximately equals $\Phi_{\rm Bn}$, and beyond this threshold the system manifests a $p$-type Schottky contact. Consequently, we successfully induce a transition from $p$-type to $n$-type Schottky contact in G/\mbip\ by applying a perpendicular positive
$E_{\rm field}$ to the system. Despite the G Fermi level shifting closer to the \mbip\ VBM, no intersection between the \mbip\ VBM and the system Fermi level is observed, even at the highest applied value ($E_{\rm field} = 0.35$~V/\AA). As a result, no formation of a $p$-type Ohmic contact occurs within the range of applied $E_{\rm field}$. 
Upon reversing the polarity of $E_{\rm field}$, the SBHs exhibit an opposite trend, increasing the separation distance between 
$\Phi_{\rm Bn}$ and $\Phi_{\rm Bp}$, thus the system remains with 
a $n$-type Schottky contact. For values below 
$E_{\rm field} = -0.4$ V/\AA\ the CBM crosses the Fermi energy, resulting in a $n$-type
ohmic contact. The nearly constant line formed by red triangles indicates that the energy gap between the VBM and CBM of the material remains unaltered in the range of applied $E_{\rm field}$.

\section{Summary and Conclusions}
In conclusion, the structural and electronic properties of the bulk and layered forms of the BiP$_3$ triphosphide have been investigated by first-principles calculations.
Phonon spectra and molecular dynamics simulations confirm that the 3D crystal of \bip~ is a metal thermodynamically stable with no bandgap. Notably, few \bip~ layers exhibit semiconductor properties only with one to four layers but behave like a metal, similar to the bulk with more than five layers. By changing the number of layers, the electronic properties of \bip~ triphosphide can be fine-tuned, making it suitable for technological applications. Based on this, the electronic properties of the G/\mbip~ van der Waals heterostructure were examined. Applying vertical strain and an external electric field can adjust the electronic properties of the G/m-\bip~heterostructure, according to our calculations. Thus, we show that few-layers \bip~ is an interesting material for realizing nanoelectronic and optoelectronic devices and is an excellent option for designing Schottky nanoelectronic devices.

\begin{acknowledgments}
The authors acknowledge financial support from the Brazilian agencies CNPq, CAPES, FAPEMIG, and INCT-Nanomateriais de Carbono, and the LCC-UFLA, CENAPAD-SP and Laborat{\'o}rio Nacional de Computa{\c{c}}{\~a}o Cient{\'i}fica (LNCC-SCAFMat2) for computer time.
\end{acknowledgments}
\bibliography{main}
\begin{appendices}
  \renewcommand\thetable{\thesection\arabic{table}}
  \renewcommand\thefigure{\thesection\arabic{figure}}
  \setcounter{figure}{0}    

\section{Lattice parameters and energy stability }
\label{label:geometries}
\begin{figure}[!htb]
    \includegraphics[width=\columnwidth]{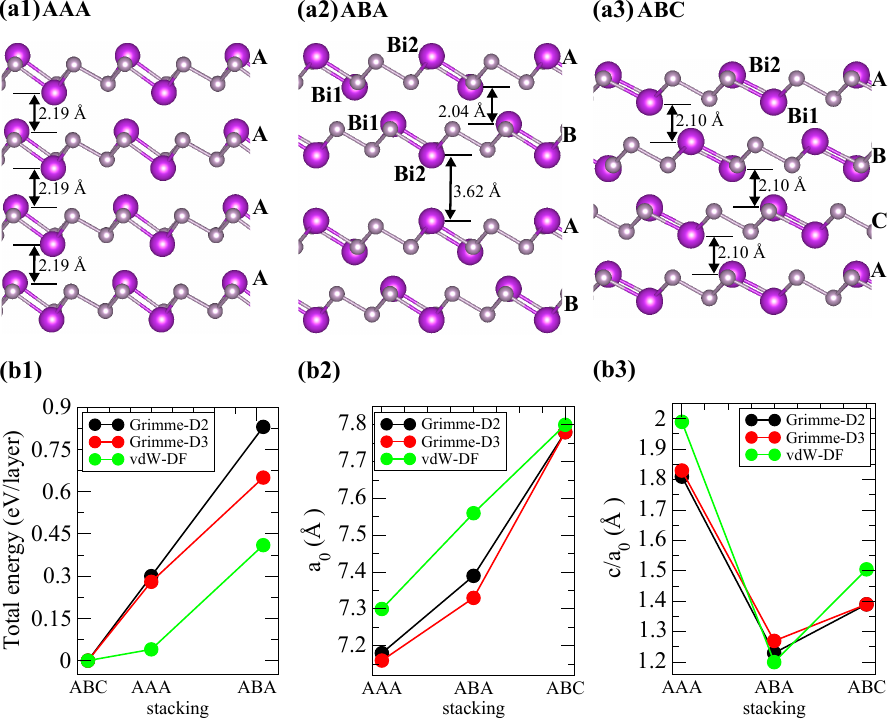}
    \caption{\label{fig:SM1} \bip-bulk structural models to the (a1) AAA-, (a2) ABA-, and (a3) ABC-stacking layers. (b1) Total energy, (b3) lattice parameter ($a_{0}$), and (b3) $c/a_{0}$ ratio to differents stacking of \bip-bulk.}
\end{figure}

We investigate three possible high symmetry stacking layers named (i) AAA-stacking; there is no shift between the layers, and the atoms of one layer are precisely aligned with those of another [Fig. \ref{fig:SM1}-(a1)]; (ii)  ABA-stacking, \bip-bulk is formed by stacking AB layers of \bip~bilayers, where one bismuth (Bi1) in each bilayer is positioned above (or below) a hexagon formed by the phosphorus of the other layer, and the remaining bismuth (Bi2) is located in the rhombohedral spaces between the AB stacked layers [Fig.\ref{fig:SM1}-(a2)],  and (iii) ABC-stacking, each layer contains bismuth (Bi1) located at the center of a hexagon formed by P atoms from the adjacent layer, and there is another bismuth (Bi2) that creates a rhombohedral space between two stacked layers [Fig.\ref{fig:SM1}-(a3)].  We also have analyzed the total energy, lattice parameter ($a_{0}$), and $c/a_{0}$ ratio for each stacking configuration [Fig. \ref{fig:SM1}-(b1)-(b3)]. The comparison of different stacking arrangements shows that AAA-stacking maintains consistent interlayer distances of 2.19\AA~within the bulk structure. However, ABA-stacking experiences a significant increase in the interlayer distance along the B-A spacing, which reaches 3.62\AA,  due to the strong repulsion between Bi atoms. In contrast, ABC-stacking shows remarkable stability with an interlayer distance of 2.10\AA, highlighting low repulsion and significant van der Waals interactions.
\section{SOC+HSE band structure of BiP$_3$ bulk}
\label{label:bulkhse}
\begin{figure}[!htb]
    \includegraphics[width=6cm]{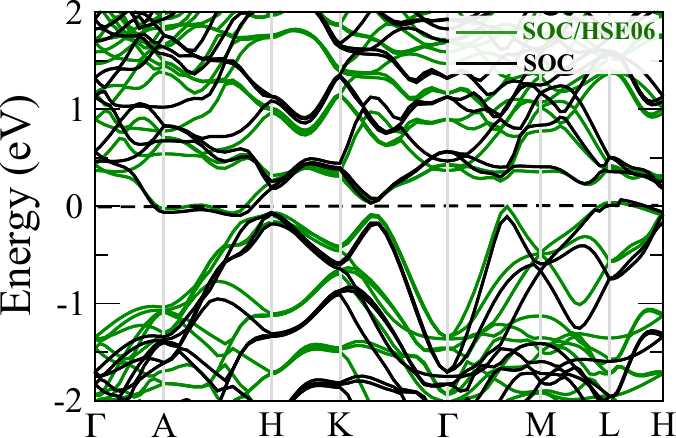}
    \caption{\label{fig:bulkhse} Band structure for \bip-bulk with SOC (black lines) and SOC+HSE06 (green lines).}
\end{figure}

This study evaluated the gap energy of \bip-bulk [Fig. \ref{fig:bulkhse}] using the Herd-Scuseria-Emzerhof hybrid functional (HSE06) calculation, employing a 10$\times$10$\times$1 k-point mesh. We observed that adding HSE06 (green lines) did not change the metallic behavior. The band structures were similar to SOC+GGA (black lines).

\section{Exfoliaton energy to mono- and bilayer of \bip}
\begin{figure}[!htb]
    \includegraphics[width=4cm]{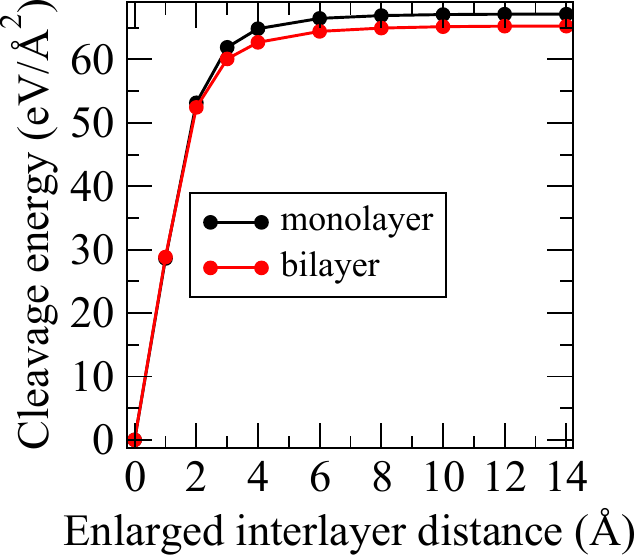}
    \caption{\label{fig:exfoliation} Exfoliation (or cleavage) energy to remove a monolayer (black line) and a bilayer (red lines) from a 6L-\bip~.}
\end{figure}

We compared the energy of a thick slab composed of six atomic layers of \bip~(6L) to that of a single (two) atomic layer far away from the 5L (4L) slab. Our aim was to calculate the exfoliation energy. We achieved this by using a large supercell that could hold 6L-\bip~while maintaining enough vacuum to prevent any interaction between its images.  To simulate the mechanical exfoliation process, We isolated one or two atomic layers from the slab. We identified the point of stabilization of total energy to determine exfoliation energy. The energy needed for exfoliating a monolayer is 67 meV/\AA$^{2}$ (black line in Fig. \ref{fig:exfoliation}), whereas for a bilayer, it is 60 meV/\AA$^{2}$ (red line in Fig. \ref{fig:exfoliation}).
\label{label:exfoliation}
\section{SOC+HSE band structure of m-BiP$_3$}
\label{label:hsemonolayer}
\begin{figure}[!htb]
    \includegraphics[width=3.5cm]{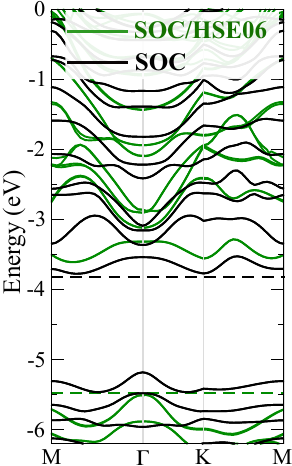}
    \caption{\label{fig:hsemonolayer} Band structure for m-bulk with SOC (black lines) and SOC+HSE06 (green lines).}
\end{figure}
\section{Structural models  of  few m-BiP$_3$ layers}
\label{label:models_2l_to_8l}
\begin{figure}[!htb]
    \includegraphics[width=7cm]{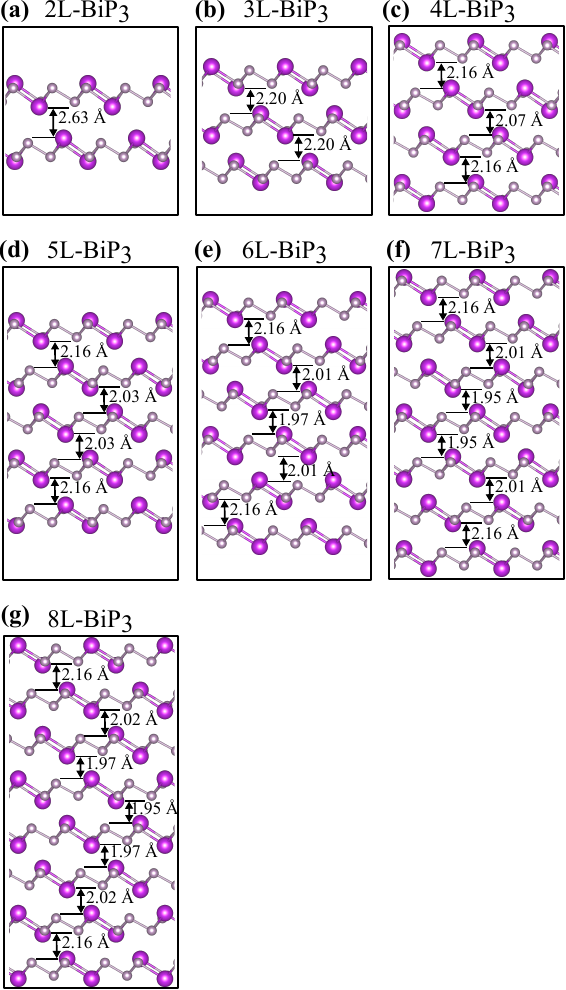}
    \caption{\label{fig:models_2l_to_8l}  Structural model for the (a) bilayer, (b) 3L, (c) 4L, (d) 5L, (e) 6L, (f) 7L, (g) 8L of \bip.}
\end{figure}
\section{Band structure of  few m-BiP$_3$ layers}
\label{label:bandfewlayers}
\begin{figure}[!htb]
    \includegraphics[width=7cm]{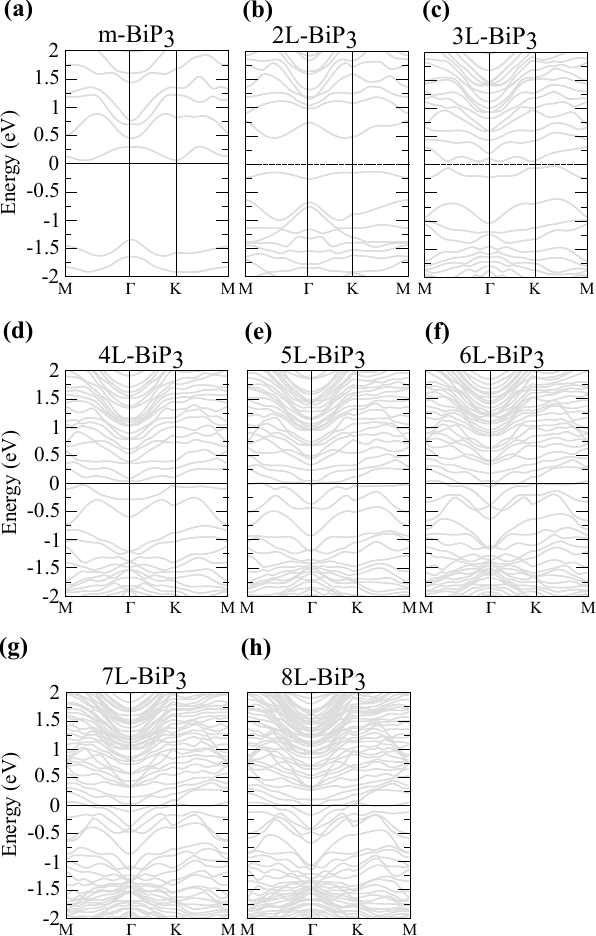}
    \caption{\label{fig:bandfewlayers}  Band structure for the (a) monolayer, (b) bilayer, (c) 3L, (d) 4L, (e) 5L, (f) 6L, (g) 7L, and (f) 8L of \bip.}
\end{figure}

\section{Graphene/\bip stacking configurations}
\label{label:configsgbip3}
\begin{figure}[!htb]
    \includegraphics[width=\columnwidth]{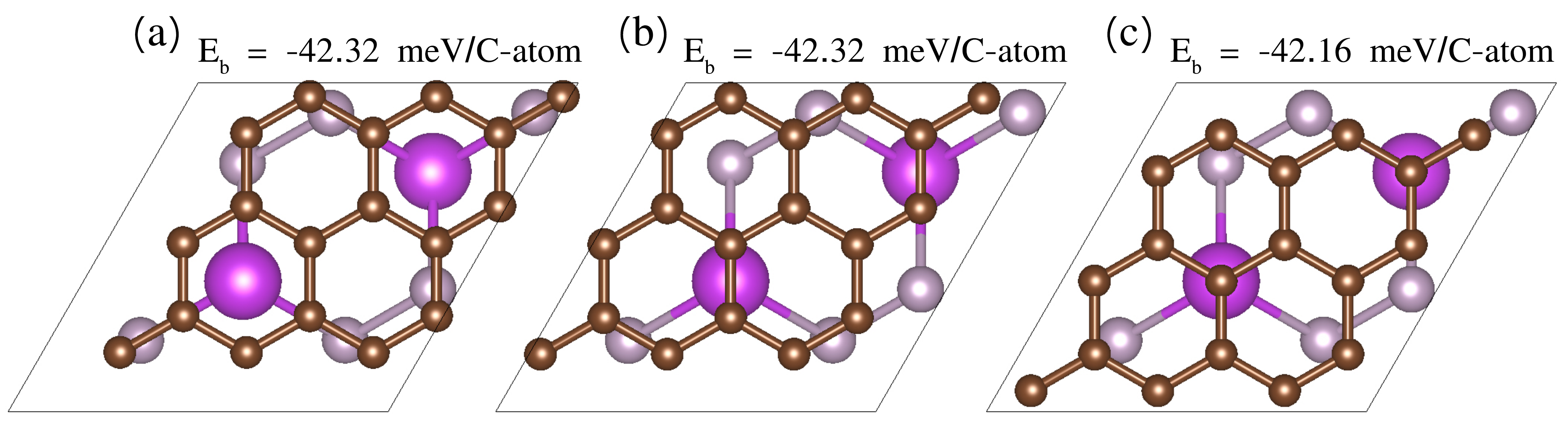}
    \caption{\label{fig:configsgbip3} The binding energy of three highly symmetric configurations of graphene on \bip~with the topmost Bi positioned in the (a) center of a carbon hexagon, on a (b) carbon bridge, and (c) aligned with graphene's carbon. }
\end{figure}

We have examined three highly symmetric configurations of the heterostructure formed by graphene on BiP3, with the topmost Bi atom placed in three distinct positions: (i) at the center of a carbon hexagon [Fig. \ref{fig:configsgbip3}-(a)], (ii) on a carbon bridge [Fig. \ref{fig:configsgbip3}-(b)], and aligned with the carbon atoms in graphene [Fig. \ref{fig:configsgbip3}-(c)]. Our analysis reveals that the most stable configuration, with the topmost Bi atom positioned in the center of a carbon hexagon, exhibits greater stability when evaluated using the vdW-DF and Grimme's DFT-D2 and DFT-D3 approaches. However, it's worth noting that the differences in stability between this configuration and the other two are marginal, with the less stable configurations showing less than a 1\% difference in the binding energy.

\end{appendices}
\end{document}